\documentclass[conference]{IEEEtran}
\IEEEoverridecommandlockouts

\usepackage{cite}
\usepackage{amsmath,amssymb,amsfonts}
\usepackage{algorithmic}
\usepackage{graphicx}
\usepackage{textcomp}
\usepackage{xcolor}
\usepackage{xspace}
\usepackage{enumitem}
\usepackage[hyphens]{url}
\usepackage{hyperref}
\def\BibTeX{{\rm B\kern-.05em{\sc i\kern-.025em b}\kern-.08em
    T\kern-.1667em\lower.7ex\hbox{E}\kern-.125emX}}

\newcommand{\mypar}[1]{\noindent \textbf{#1:}}

{

\newcommand{\eg}{\textit{e.g.}}

\newcommand{\ignore}[1]{}

\newcommand{\revisionhighlight}[1]{#1}

\newcommand\douppercase[1]{\ifnum\ifhmode\spacefactor\else2000\fi>1000 \uppercase{#1}\else#1\fi}

\newcommand{\ramp}{RAMP\xspace}
\newcommand{\rampdm}{RAMP-DM\xspace}

\setlength{\abovedisplayskip}{3pt}
\setlength{\belowdisplayskip}{3pt}

\newcommand{\pbdue}{{p_{pb,due}}}
\newcommand{\pbnde}{{p_{pb,nde}}}
\newcommand{\pldue}{{p_{lb,due}}}
\newcommand{\plnde}{{p_{lb,nde}}}
\newcommand{\pcdue}{{p_{c,due}}}
\newcommand{\pcnde}{{p_{c,nde}}}

\begin{document}

\title{Analyzing a Two-Tier Disaggregated Memory Protection Scheme Based on Memory Replication}\vspace{-1cm}


\author{
    \IEEEauthorblockN{Haris Volos}
    \IEEEauthorblockA{\textit{University of Cyprus} \\
    hvolos01@ucy.ac.cy}
    \and
    \IEEEauthorblockN{Yiannnakis Sazeides}
    \IEEEauthorblockA{\textit{University of Cyprus} \\
    yanos@ucy.ac.cy}
}

\maketitle

\thispagestyle{plain}
\pagestyle{plain}

\begin{abstract}
As memory technologies continue to shrink and memory error rates increase, the demand for stronger reliability becomes increasingly critical.
Fine-grain memory replication has emerged as an appealing approach to improving memory fault tolerance by augmenting conventional memory protection based on error-correcting codes with an additional layer of redundancy that replicates data across independent failure domains, such as replicating memory pages across different NUMA sockets.
This method can tolerate a broad spectrum of memory errors, from individual memory cell failures to more complex memory controller failures.
However, applying memory replication without a holistic consideration of the interaction between error-correcting codes and replication can result in redundant duplication and unnecessary storage overhead.
We propose Replication-Aware Memory-error Protection (RAMP), a model that helps explore error protection strategies to improve the storage efficiency of memory protection in memory systems that utilize memory replication for performance and availability.
We use RAMP to determine a protection strategy that can lower the storage cost of individual replicas while still ensuring robust protection through the collective protection conferred by multiple replicas.
Our evaluation shows that a solution derived with RAMP enhances the storage efficiency of a state-of-the-art memory protection mechanism when paired with rack-level replication for disaggregated memory.
Specifically, we can reduce the storage cost of memory protection from 27\% down to 17.7\% with minimal performance overhead. 
\end{abstract}


\section{Introduction}
\label{sec:introduction}
System reliability is a critical factor in the design and operation of modern computing systems, from IoT devices~\cite{xing:iot-reliability:ieee-iotj:2020, philip:iot-healthcare:jsac:2021} to high-performance computing~\cite{schroeder:hpc-failures:tdsc:2010} and data-center infrastructure~\cite{beyer:sre-google:oreilly:2016}. 
Memory reliability plays a key role in this context, as memory errors, ranging from bit flips to complete module failures, can severely impact system reliability, leading to crashes, data corruption, and significant performance degradation~\cite{meza:dramfailures:dsn:2015}. 
As memory densities continue to increase~\cite{lee:ddr5:iedm:2023, zhang:pm-chipkill:micro:2018}, and as system architectures become more complex, the frequency of memory errors is also on the rise, making addressing memory reliability critically important~\cite{beigi:dram-faults:hpca:2023}.

Fine-grain memory replication is emerging as a promising approach to improving memory reliability. It augments per-DIMM ECC with an additional layer of redundancy, ensuring that critical data has multiple copies across independent
failure domains, such as different memory channels~\cite{zheng:raim:isca:2017} or memory controllers~ \cite{patil:dve:isca:2021}. 
This approach not only mitigates errors within a single module, such as single-bit errors, but also increases robustness against errors that extend beyond the module, including those caused by memory channel or memory controller failures~\cite{meza:dramfailures:dsn:2015}.
It is particularly useful in Non-Uniform Memory Access (NUMA) systems, where coherent replication can improve both reliability and performance~\cite{patil:dve:isca:2021}. 
In disaggregated memory systems, where memory is separated from processing units, techniques such as replication and erasure coding are employed to ensure fault tolerance by encoding data into multiple fragments, allowing recovery even if part of the memory fails~\cite{lee:hydra:fast:2022, zhou:carbink:osdi:2022, tsai:dpm:atc:2020, shan:legoos:osdi:2018}. 

Introducing redundancy at different levels of the memory hierarchy, from per-DIMM ECC to memory replicas, without a holistic approach can result in unnecessary storage overheads due to duplication. 
A two-tier memory resilience scheme that decouples error detection and correction, using error detection in the first tier and replication for error correction in the second tier, can help mitigate this issue~\cite{patil:dve:isca:2021}. 
However, several approaches apply replication without considering the first tier~\cite{lee:hydra:fast:2022, zhou:carbink:osdi:2022, tsai:dpm:atc:2020, shan:legoos:osdi:2018}, potentially leading to excess duplication, overhead, and reduced efficiency. 
We believe this problem arises from the absence of a unified model to reason about protection, storage efficiency, and performance interactions between the two tiers.

We propose \emph{\textbf{R}eplication-\textbf{A}ware \textbf{M}emory-error \textbf{P}rotection} (\ramp) to fill this gap.
\ramp seeks to offer a framework for designing and analyzing two-tier memory resilience schemes, where memory replication is used to handle errors that the first tier cannot tolerate. 
\ramp provides analytical models that enable system designers and operators to understand the interaction between the two protection tiers and assess how the first tier's protection strength affects the overall protection provided by multiple replicas in the second tier. 
This guidance helps determine the optimal protection strength for individual replicas in the first tier and appropriate replication level in the second tier, resulting in improved efficiency.
For example, RAMP enables the evaluation of using weaker, more efficient, and lower-overhead ECC at individual memory modules while relying on replicated data in other modules to provide collective protection and tolerate memory errors.

We demonstrate the utility of our approach by applying it to a state-of-the-art rack-level resilience mechanism designed for disaggregated memory systems~\cite{lee:hydra:fast:2022}, combined with a recent chipkill-correct design for high-density non-volatile memory~\cite{zhang:pm-chipkill:micro:2018}.
This combination tolerates memory errors through a two-tier cross-layer resilience scheme: 
(i) the chipkill-correct mechanism provides a memory-protection tier that uses the memory chip failure protection bits to detect and opportunistically correct memory errors at high performance, and 
(ii) the rack-level resilience mechanism provides a memory-replication tier that uses rack-scale replication and erasure coding to correct memory errors that are detected but uncorrected by the memory-protection tier at low storage cost. 
Using our analytical framework, we show how weakening the chipkill protection of each individual replica, we can reduce storage cost from $27\%$ down to $17.7\%$ while we attain the same protection level as the original design through the collective protection of multiple replicas, with minimal performance overhead. 

In particular, we make the following contributions:

\begin{itemize}[leftmargin=*,noitemsep,topsep=0pt]

\item We develop \ramp, an analytical framework to reason about the protection and storage trade-offs in two-tier memory resilience schemes. 

\item We use our framework to design \rampdm, a two-tier memory resilience scheme for disaggregated memory that combines memory protection and replication to tolerate memory errors.

\item We evaluate reliability, efficiency, and performance trade-offs in \rampdm, showing that our framework can help improve storage costs with minimal performance overhead.

\end{itemize}
\section{Background \& Motivation}
\label{sec:background}

\subsection{Memory Technologies}
DRAM is the dominant type of memory technology used for main memory in both server and client systems. 
Each DRAM cell stores a bit as a capacitor charge, accessed by a transistor. 
However, as DRAM technology scales to smaller cell geometries, it encounters significant challenges related to reliability~\cite{lee:ddr5:iedm:2023, patil:dve:isca:2021}.

Non-volatile memory (NVM) technologies, such as phase-change memory (PCM), spin-transfer torque RAM (STT-RAM), and resistive random access memory (ReRAM), are emerging as viable alternatives to DRAM for main memory due to their higher storage density and competitive performance~\cite{lee:pcm:isca:2009, kultursay:sttram:ispass:2013, xu:reram:hpca:2015}. 
For example, the now discontinued Intel Optane DC Persistent Memory (DCPMM) DCPMM offered increased capacity with 300ns read latency ~\cite{yang:optane-guide:fast:2020}, while Weebit ReRAM offers 20ns read latency \cite{weebit:skywater-ip}.


\subsection{Memory Systems}

\begin{figure}[!t]
\centering
\includegraphics[width=3.2in]{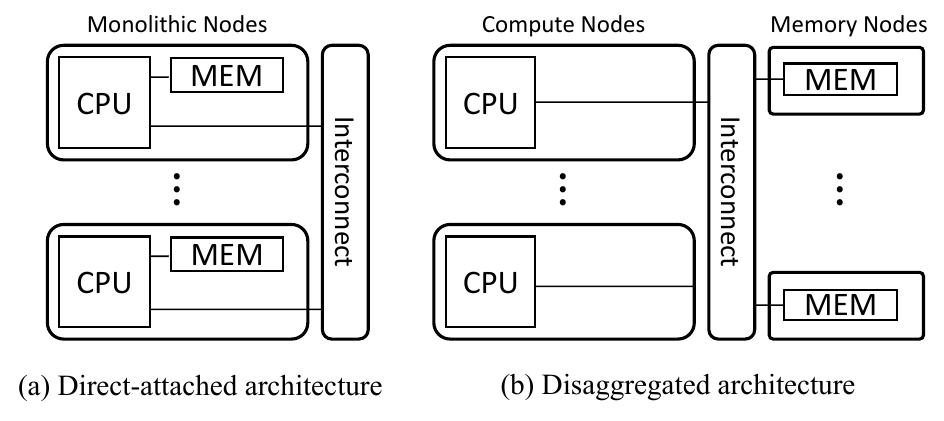}
\caption{Memory system architectures}
\label{fig:memory-system-architectures}
\vspace{-0.5cm}
\end{figure}

DRAM server memory systems organize memory bit cells into arrays, which are grouped into chips, and further assembled into DIMM (Dual Inline Memory Module) modules. 
These modules are accessed by the memory controller through one or more channels, which coordinates data transfer between the processor and the memory system.
Data is transferred in memory blocks, with each block typically 64 bytes in size.
Although the microarchitecture design of NVM subsystems is more complex than conventional DRAM systems, at a high level, NVM systems follow similar chip structure and system organization as DRAM systems~\cite{yang:optane-guide:fast:2020, wang:model-nvm:micro:2020}.

Conventional Non-Uniform Memory Access (NUMA) systems consist of multiple sockets, each containing a processor with memory directly attached. 
Scaling up compute and memory capacity by adding more sockets is prohibitively expensive~\cite{barroso:wsc:book:2018, lim:memory-blade:isca:2009}, so datacenter systems typically scale out by adding additional server nodes, as shown in Figure~\ref{fig:memory-system-architectures}(a). 
However, since applications often exhibit imbalanced memory usage across server nodes, this can lead to underutilization of the aggregate memory, resulting in poor memory efficiency~\cite{lu:pond:asplos:2023, gu:infiniswap:nsdi:2017}.

Disaggregated memory architectures decouple the processor from memory into separate compute and memory nodes, which are interconnected by a high-performance system interconnect, as shown in  Figure \ref{fig:memory-system-architectures}(b)). 
Compute nodes mainly provide processing capability, but they may also include a small amount of local memory used as a local cache.
Memory nodes provide memory capacity by attaching standard memory subsystems to the network, either through an I/O interface, such as Remote Direct Memory Access (RDMA)~\cite{rdma-consortium}, or a memory semantics interface, such as Compute Express Link (CXL)~\cite{cxl-consortium}.
Overall, memory disaggregation allows separate evolution and scaling of processing and memory, which lets tailoring the compute-to-memory ratio to the specific needs of the workload, and improves memory utilization as memory is shared across multiple compute nodes.

\subsection{Memory Errors}
\label{sec:failure-model}


Data stored in a memory system is susceptible to memory errors, which can be classified into two main categories. 

\mypar{Cell Errors}
Memory cells are susceptible to both transient and permanent device faults.
DRAM cells have been traditionally susceptible to transient faults, such as radiation-induced errors~\cite{schroeder:dramfailures:sigmetrics:2009}. However, DRAM is increasingly vulnerable to permanent device faults due to manufacturing variability and defects arising from miniaturization~\cite{cha:dram-defects:hpca:2017}, as well as wear-out faults caused by aging~\cite{fieback:dram-aging:delft:2017}.
NVM cells can experience permanent faults due to limited and variable endurance, as well as transient faults caused by resistance drift and read disturb~\cite{yoon:freep:hpca:2011}.
These NVM cell errors are random in nature~\cite{zhang:pm-chipkill:micro:2018}. 
Raw bit error rate (RBER) in PCM and ReRAM is significantly higher than in DRAM and ranges from $10^{-3}$ to $10^{-5}$~\cite{zhang:pm-chipkill:micro:2018}, depending on the technology and time since last write or refresh.

\mypar{Non-Cell Errors}
Memory errors can also occur when other components of the memory system fail, including memory controller and memory channel failures. 
These can be caused by transient faults due to signal disturbances, as well as permanent failures resulting from faults in the logic and transmission circuitry~\cite{meza:dramfailures:dsn:2015}.

\subsection{Memory Error Protection}

To protect against memory errors, memory systems maintain error correcting codes (ECC) computed over data. 
These codes can detect and correct a small number of errors.
For example, single error correction double error detection (SEC-DEC) uses parity to detect up to two-bit errors or correct a single-bit error.
Chipkill uses wider ECC to protect against multi-bit errors and chip failures.
Detectable but uncorrectable memory errors (DUE), which are detected but cannot be corrected by ECC, can cause memory system failures.
Non-detectable memory errors (NDE), which are non-detected and potentially miscorrected by ECC, do not cause memory system failures but may cause silent data corruption (SDC), which is also higly undesirable. 

As memory technology continues to shrink and density increases, ECC storage overheads required to maintain reliability are growing. 
High-density DDR5 DRAM chips, for example, include on-die ECC, and DDR5 DIMMs double the number of error-correcting bits compared to DDR4 DIMMs~\cite{micron:ddr5:whitepaper}
Similarly, for dense NVM with high RBER, using stronger codes to achieve a low uncorrectable bit error rate (UBER) and low SDC rate incurs prohibitive storage overheads~\cite{zhang:pm-chipkill:micro:2018}.
These overheads remain significant ($\sim27\%$) despite recent efforts on improving storage
efficiency~\cite{zhang:pm-chipkill:micro:2018}. 
Additionally, stronger ECC may not be sufficient for addressing errors that occur when both ECC and ECC-protected data are colocated within the same failure domain, such as within a single memory channel or memory controller.

\begin{figure}[!t]
\centering
\includegraphics[width=2.5in]{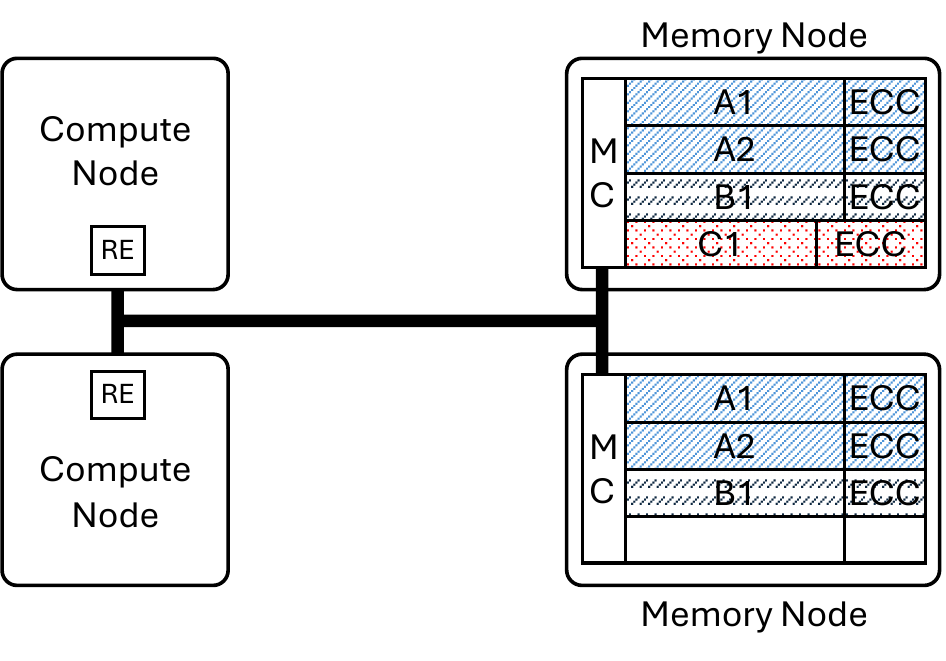}
\caption{Replication-Aware Memory-error Protection.}
\label{fig:ramp-architecture}
\end{figure}

Fine-grain memory replication is emerging as an approach
to improving memory reliability by augmenting per-DIMM ECC with an additional layer of redundancy, ensuring that critical data has multiple copies across independent failure domains, such as different memory channels~\cite{zheng:raim:isca:2017} or memory controllers~ \cite{patil:dve:isca:2021}.
For example, Figure~\ref{fig:ramp-architecture} shows three applications A, B, and C with different degrees of replication.
Application A and B have two replicas per page so they can tolerate memory controller failures.

\ignore{
\subsection{Memory errors and their handling}
\label{sec:failure-model}
Disaggregation provides separate fault domains between processing and memory, meaning that the failure of a compute node does not render disaggregated memory unavailable, and vice versa, that is when a memory node fails, compute and other memory nodes continue to function.

In this work, we focus on memory node failures. These may happen due to several reasons. First, a memory node may fail due to a random NVM bit cell error. Bit cells are susceptible to permanent failures due to limited and variable endurance, and transient failures due to resistance drift and read disturb~\cite{yoon:freep:hpca:2011}. RBER in PCM and ReRAM is significantly higher than in DRAM, ranging from $10^{-3}$ to $10^{-5}$~\cite{zhang:pm-chipkill:micro:2018}.
Second, a memory node may fail due to a memory subsystem failure. 
At a very high level, NVM subsystems follow similar chip structure and system organization as DRAM subsystems, comprising a memory controller that is connected to multiple memory chips through one or more channels.
However, their microarchitecture design is more complex than conventional DRAM subsystems~\cite{wang:model-nvm:micro:2020}.
Hence, NVM subsystems will likely suffer from similar or more complex failures, including transient failures due to signal disturbances, and permanent failures due to faults in logic and transmission circuitry. 
Finally, a memory node may fail due to 
failure in the network-interface card (NIC) that connects a memory node with the rest of the system.
}

\ignore{
Memory nodes incorporate hardware-level error-correcting code (ECC) mechanisms to protect against memory errors. Memory nodes fail when such protection mechanisms detect but cannot correct an underlying memory error. 

Tolerating NVM cell failures involves device- and architecture-level techniques that mitigate endurance-related permanent failures through write-efficient coding, memory remapping  and embedded redirection of failed lines8, , and mitigate transient failures through ECC8. However, maintaining a correctable error rate below 10-15 that is necessary with petabyte memory sizes expected in rack-scale DM requires using strong BCH ECC with high energy and die area overheads, up to 30

Silent data corruption?

memory node failures can happen due to: cell, chip, controller, nic, failures
hardware-level protection techniques addrsss cell, chip failures
software-level replication addresses uncorrected errors

technology: persistent memory, nvram, 
failure model:
- disaggregation failures: compute nodes, memory nodes: memory controllers, bit errors, 
- memory bit errors: cite rber of pcm/reram

compute and memory nodes fail independently

How do memory nodes fail?

Bit errors in NVM, memory controller errors, circuit errors (how do these differ from bit errors?)
- sources of bit errors: nvm cells, circuit lines, etc. Here, we focus on nvm cells

ECC addresses bit errors

assumption: 
-memory nodes fail similarly to servers today?
-applications will rely on replication and erasure coding to recover from errors that are not recoverable with ECC 
-investigate bit errors as the primary failure model; won't model impact of memory controller and circuit failures. most likely, that will follow dram circuits. leave this as future work.

Hardware-level protection techniques, such as parity and chipkill-correct, employ redundancy to guard against bit corruption due to memory cell and/or chip failures. 

However, memory nodes can still fail due to memory controller

Memory nodes can fail either due to 

why replication for performance? To avoid a single memory node becoming a performance bottleneck
why replication for availability? To tolerate failures that render the complete memory node unavailable and which cannot be addressed by bit and chip protection techniques. This includes memory controller failures, power delivery/supply?, NIC failures? 

Explore co-design of memory protection techniques for such applications
}
\section{Replication-Aware Memory-error Protection}
\label{sec:shepherd}

We propose \emph{\textbf{R}eplication-\textbf{A}ware \textbf{M}emory-error \textbf{P}rotection} (\ramp), a framework for designing and tuning computing systems employing memory replication with memory protection to tolerate memory errors.


\subsection{System Model}
Figure~\ref{fig:ramp-architecture} presents an abstract system model of computing systems that employ memory replication together with memory protection.
This model captures a variety of systems, ranging from conventional NUMA systems~\cite{patil:dve:isca:2021} to emerging disaggregated memory architectures~\cite{lee:hydra:fast:2022, zhou:carbink:osdi:2022, tsai:dpm:atc:2020, shan:legoos:osdi:2018}.
In this model, compute nodes provide processing power, while memory nodes offer memory capacity and implement memory protection (e.g., ECC) for their memory. 
Although memory nodes are shown as separate from compute nodes in this abstract model, the actual implementation may integrate them, as seen in NUMA systems, where a NUMA node consists of both processors and direct-attached memory~\cite{patil:dve:isca:2021}. Compute nodes interact with memory nodes through a replication engine (RE), which implements the replication logic and coordinates data replication across multiple memory nodes.

\subsection{Tolerating Memory Errors}
\ramp assumes a computing system tolerates memory errors through a two-tier protection scheme. 
The first tier, or memory-protection tier, is a performance-optimized tier that reuses the memory chip failure protection bits (\eg, ECC) to detect and opportunistically correct memory errors at high performance, while ensuring that no miscorrection occurs.
The second tier, or memory-replication tier, is a storage-optimized tier that employs memory replication across multiple memory nodes to correct memory errors that are uncorrected by the memory-protection tier, including errors due to cell failures, chip failures, channel failures, and complete memory node failures

The memory-protection tier provides configurable protection against memory errors.
It requires that the memory controller be able to compute and decode different codes, as well as a mechanism to determine which ECC technique to use for each memory access. 
For example, the memory-protection tier could support page-granularity protection by augmenting the virtual memory page table and TLB to include protection information for each page~\cite{yoon:virtualized-ecc:asplos:2010}. 
The configuration capability can range from allowing upper tiers select a memory protection scheme from a fixed set of protection schemes (\eg, SEC-DEC, chipkill) to providing complete flexibility in choosing the protection method and strength~\cite{yoon:virtualized-ecc:asplos:2010}.

Memory nodes that fail correction at the memory-protection tier report DUEs to the replication engine for further handling. 
Memory nodes can report DUEs either by piggybacking on coherence requests~\cite{patil:dve:isca:2021} or by leveraging hardware error reporting mechanisms, such as Intel Machine Check Architecture (MCA)~\cite{intel:mce:book:2024, xu:nova-fortis:sosp:2017}, to raise a machine check exception (MCE) in response to uncorrectable memory errors.
After reporting, a memory node remains operational and continues to serve memory accesses to non-failed
memory regions, thus improving availability. 
Depending on the overhead of the error reporting mechanism, the ability to opportunistically correct errors in the memory-protection tier, in addition to detection, becomes crucial for preventing severe performance slowdowns due to error reporting~\cite{meza:dramfailures:dsn:2015, barroso:wsc:book:2018}.

For each replicated data item, the memory-replication tier maintains multiple replicas across memory nodes. 
The memory-replication tier maps each replica to a memory node and memory region, and configures the hardware protection strength of each replica to meet a target UBER and SDC rate, following application requirements. 
The memory-replication tier may track and blacklist failed memory regions to avoid mapping replicas to regions with known errors. 
When the memory-replication tier accessing a data item faces a DUE, it attempts to correct the memory error using another replica.

The memory-replication tier can implement any block-level static homogeneous replication method, including primary-backup replication, chain replication, quorum-based replication, and erasure coding. 
Static requires a fixed number of replicas whose protection strength does not change dynamically, thus relieving the replication engine from having to support frequent protection changes.
Homogeneous requires all replicas to have the same protection strength, thus simplifying replica strength reasoning.

\ignore{
    A lightweight service processor on the memory node handles the exception and returns an error as a response to the RDMA request by piggybacking on the existing error reporting mechanism of RDMA.
}
\ignore{
    To serve control plane operations and support fine-grain error reporting, the memory nodes include a lightweight service processor.
}

\ignore{
On read failure, it redirects the request to another replica.
On write failure, if the memory node remains operational, then it may attempt to write to another memory region within the same memory node. Otherwise it the memory node fails completely, software issues the writes to another memory node, and also remaps/migrates (asynchronously) all other replicas of the failed memory node. 
}

\ignore{
operation:
-compute nodes access memory using RDMA; rdma offers robust failure semantics compared to ld/st; piggyback on existing error reporting mechanism
-when an RDMA causes a memory side error, memory nodes do not crash, report operation failure, compute nodes recover by trying another replica, memory nodes remain functional
-memory nodes do not crash; instead poison affected region and continue servicing other requests (rely on an extended from of intel machine check architecture); current mce raises exception on read errors only. report both reads and non-posted-writes. we expect replication protocols to use non-posted writes to ensure writes reach nvm (cite snia)
}

\ignore{
Application software running on compute nodes may tolerate memory errors using application-level redundancy in the form of replication and checksumming.
\ramp does not dictate or implement a specific redundancy scheme, leaving the implementation to the application for maximum flexibility. 
Because targeted applications already employ redundancy for performance and availability (\cref{sec:ramp:idea}), we do not expect this requirement to significantly burden applications. 

Applications may use replication to tolerate DUEs. 
For each replicated data item, an application maintains multiple replicas across memory nodes.
Applications map each replica to a memory node and memory region, and configure the hardware protection strength of each replica to meet a target UBER and SDC rate.
Applications may also track and blacklist failed memory regions to avoid mapping replicas to regions with known errors.
When an application trying to access a data item faces a DUE, it attempts to correct the memory error using another replica. 

Applications may use checksumming to tolerate NDEs, including non-detectable bit cell errors and scribbles, that would otherwise silently corrupt data.
With checksumming, an application maintains a checksum for each data item.
When the application writes a data item, it computes and stores a corresponding checksum. When the application later reads the data, it may recompute the checksum and verify that the computed checksum matches the stored checksum.
}
\ignore{Application-level checksumming increases CPU utilization, but provides end-to-end protection against silent data corruption.}

\ignore{
Software running on a compute node accesses disaggregated memory using one-sided remote DMA (RDMA) reads and writes. 
When the network interface card (NIC) at a memory node receives an RDMA request, it performs a local DMA request to the node's memory controller, which in turn issues memory accesses to memory media.
The controller uses hardware ECC to detect and correct memory errors, and leverages existing hardware error reporting mechanisms, such as Intel Machine Check Architecture (MCA), to report DUEs. 
The controller transparently corrects correctable errors, and silently returns invalid data for undetectable errors.
For DUEs, the controller raises a hardware exception in response to uncorrectable memory errors. 
A lightweight service processor on the memory node handles the exception and returns an error as a response to the RDMA request by piggybacking on the existing error reporting mechanism of RDMA. 
After reporting the error, the memory node continues normal operation by servicing other pending RDMA requests. 
For error reporting mechanisms, that do not provide a mechanism for detecting store failures, like Intel Machine Check Architecture (MCA), the NIC issues an additional read after a write to check success of the write. 
}

Overall, \ramp enables co-designing memory replication together with memory protection to trade-off protection strength, storage efficiency, and performance.
\ignore{Maintaining multiple replicas across memory nodes enables the power of many choices. Instead of trying hard to eliminate uncorrectable memory errors within a single memory node using strong but expensive codes, \ramp accepts the possibility of uncorrectable errors.}
Computing systems that maintain multiple replicas across memory nodes can employ weaker but lower-storage-overhead ECC within individual replicas. 
While weaker ECC increases failure rate of individual replicas, a memory system can rely on the multiple choices offered by available replicas in other memory nodes to correct a memory error. 
For example, in Figure~\ref{fig:ramp-architecture}, applications A and B have two replicas per page, so we can use weaker ECC, relying on the collective protection of multiple replicas to tolerate the increased per-replica error rate.


\subsection{Choosing Replica Protection Strength}

A key challenge in applying \ramp is choosing the right hardware protection strength of individual replicas. 
Weakening hardware-level protection of individual replicas lowers storage cost but makes DUEs and NDEs more frequent, increasing UBER and SDC rates respectively.
We can recoup the lost UBER by correcting a DUE using available replicas, at the expense of a performance overhead to access and process additional replicas.
However, we cannot always rely on replicas to recoup the lost SDC rate. 
This is because a NDE that silently corrupts data may not trigger a correction by the storage-optimizer tier, unless the application can detect the error through other means, such as checksumming~\cite{zhang:pangolin:atc:2019}.
Hence, SDC rate may limit how much we can weaken individual replica strength.

To help application and system designers choose protection strength, we develop an analytical model that estimates the expected reliability and expected performance overhead when using available replicas to correct a DUE. 
For the reliability, we estimate the combined DUE resulting from using replicas to correct a memory error. 
Because NDE does not benefit from replication, we do not compute a combined NDE. 
However, we do estimate the NDE of each individual replica as a lower reliability bound. 
For the performance overhead, we compute the average number of additional replicas that are read to correct a memory error.
We assume that reading and processing each replica contributes fixed network bandwidth and CPU overhead per replica.

Our analytical model targets block-level replication, which is a common replication approach. The model differentiates between logical and physical blocks. The logical block is the unit of recovery, that is the smallest unit of data that can be recovered by the replication protocol. To enable recovery, a replication protocol maps a logical block to multiple physical block replicas stored across multiple memory nodes. Reading a logical block may involve reading one or more physical blocks.

\ignore{
Since we focus on protection techniques against random bit cell errors, our model focuses on read failures caused by uncorrectable memory errors due to random bit cell errors. 
Our model ignores other correlated failures that affect multiple bits and blocks, such as chip or channel failures due to logic circuit errors.
}

\ignore{
The model uses the following parameters: CPU cache-line size: $c$:, Physical-block size: $b$, Cache-line failure probability due to DUE: $p_c$, Physical-block failure probability due to DUE: $p_b$.
}

\begin{table}
\caption{Analytical model symbol notation}
\label{tab:model}
\centering
\begin{tabular}{lp{6.5cm}}
\textbf{Symbol(s)} & \textbf{Description} \\
$c$, $b$        & \scriptsize{Cache-line size and physical-block size}\\
$\pcdue$, $\pcnde$  & \scriptsize{Cache-line failure probability due to DUE and NDE}\\
$\pbdue$, $\pbnde$  & \scriptsize{Physical-block failure probability due to DUE and NDE}\\
$\pldue$  & \scriptsize{Logical-block failure probability due to DUE}\\
\end{tabular}
\end{table}

The model uses the symbol notation shown in Table \ref{tab:model}.
Cache-line failure probability is the probability to fail when reading a cache-line worth of data from the memory system.
This probability depends on the memory protection scheme and the RBER of the underlying memory technology.
Although the RBER may vary over the lifetime of a memory technology, for the sake of simplicity, we adopt a single worst-case value based on the RBER at the end of a specified time period. 
For example, in NVM where the RBER may increase with the amount of time since the last write or refresh~\cite{zhang:pm-chipkill:micro:2018}\ignore{, which can range from a week to a year}, we can use the RBER at the end of the refresh period as the worst-case value.

Physical-block failure probability is the probability to fail when reading a physical-block worth of data from the memory system.
Successfully reading a physical block entails successfully reading all the cache lines that comprise the block. 
We can derive the physical-block failure probability as follows:
\ignore{from the cache-line failure probability:}


\[
\pbdue= 1 - (1-\pcdue)^{b/c}
\]
\[
\pbnde= 1 - (1-\pcnde)^{b/c}
\]

We next provide analytical models for replication and erasure coding.

\mypar{Replication}
For each logical block, the replication protocol maintains a primary physical block and a sequence of N-1 backup physical replica blocks, with the physical block size equal to the logical block size. 

When a compute node needs to read a logical block, it first reads the primary physical block. If the read fails because of a DUE, then it tries the next backup physical block in the sequence, continuing this
process until it successfully reads a block. 
When the compute node exhausts trying all available physical blocks without successfully reading one, the logical-block read fails with a DUE.

Hence, the probability to have a DUE when reading a logical block is the joint probability of all physical-block reads to fail:

\[
\pldue = \pbdue^{N}
\]

The probability to have a NDE when reading a logical block is the union probability of any physical-block reads to fail due to NDE:

\[
\plnde = \sum_{i=0}^{N-1} (1-\pbdue)^i \pbnde
\]

The average number of additional physical blocks that are read after failing to read the first physical block is:

\[
\begin{split}
a_r&= -1 + \sum_{i=0}^{N-1} \pbdue^i(1-\pbdue)(i+1)
\end{split}
\]


\mypar{Erasure Coding}
Erasure coding provides redundancy without the overhead of complete replication.
Erasure coding divides a logical block into K physical blocks,
and then encodes these blocks using Reed-Solomon code RS(K, N)~\cite{reed:code:journal-applied-math:1960} to generate R parity blocks, for a total of N=K+R blocks.
It finally writes each of these N blocks to a different remote memory node. 
Physical block size is equal to $\frac{\text{logical block size}}{K}$.

When a compute node needs to read a logical block, it can perform the read using any K physical blocks of the N physical blocks. If the compute node fails to read any of the physical blocks (due to a DUE), then it tries another physical block. 
When the compute nodes exhausts trying all available physical blocks without successfully reading K blocks, the logical-block read fails with a DUE.

Hence, the probability to have a DUE when reading a logical block is the probability to have at least N-K+1 physical blocks fail due to a DUE:


\begin{equation*}
\pldue = \sum_{i=N-K+1}^{N}f(i,N,\pbdue)
\end{equation*}

\begin{equation*}
\mathrm{where}\quad
f(k,n,p) = \binom{n}{k}p^{k}(1-p)^{n-k}
\end{equation*}

The probability to have a NDE is the number of ways in which we can read K physical blocks plus extra physical blocks due to DUE multiplied by the probability to have physical blocks fail due to DUE multiplied by the probability to have at least one physical block fail due to a NDE: 

\begin{multline*}
\plnde = \sum_{i=0}^{N-K}\binom{N}{K+i}f(i,K+i-1,\pbdue) \\ 
\times \sum_{j=1}^{K-1}f(j,K-1,\pbnde)
\end{multline*}

The average number of additional physical blocks that are read after failing to read any of the first K physical blocks is:


\begin{equation*}
a_r= -K + \sum_{i=0}^{N-K} \binom{N}{K+i} f(i, K+i-1, \pbdue) (K+i)
\end{equation*}

\noindent where each sum term is the number of ways in which we can read physical blocks multiplied by the probability to have physical blocks fail due to DUE multiplied by the number of physical blocks read.

\section{\rampdm: Resilient Disaggregated Memory}

\begin{figure}[!t]
\centering
\includegraphics[width=2.5in]{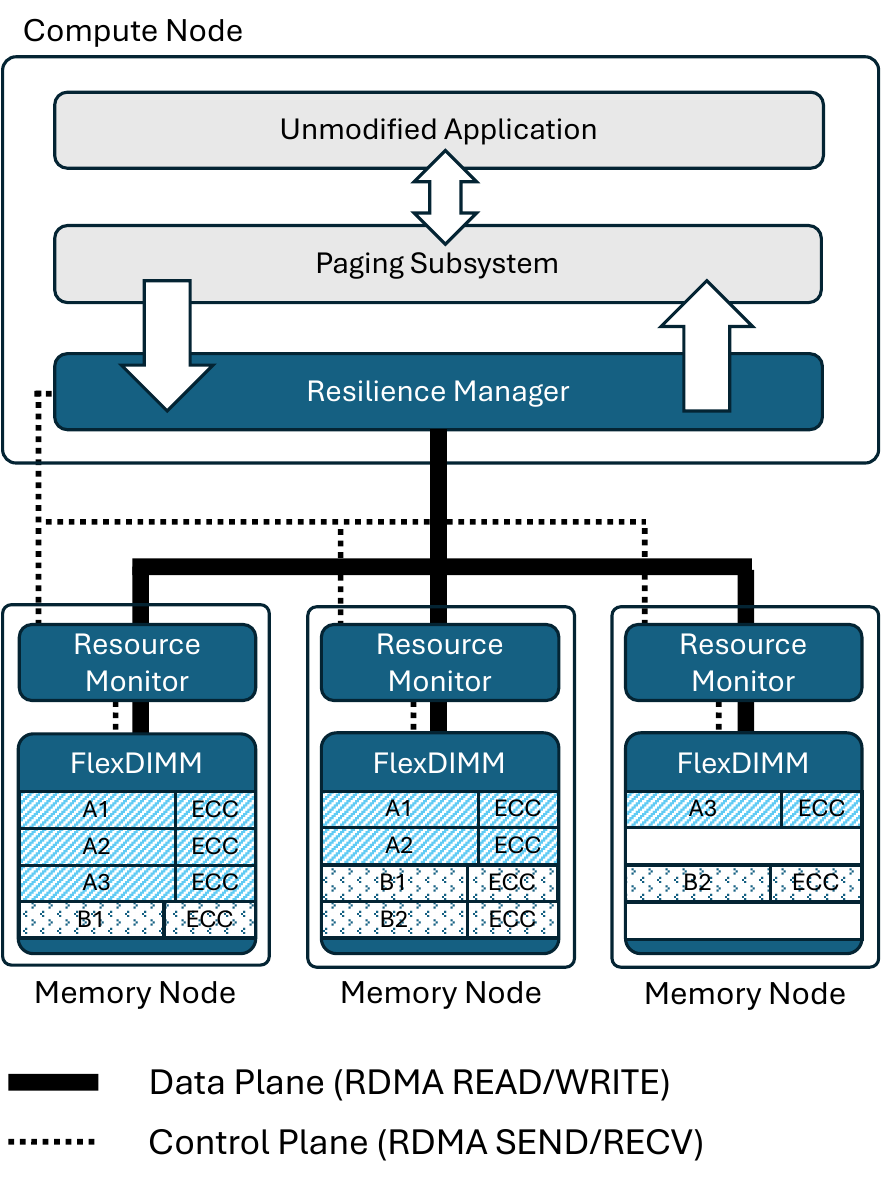}
\caption{\rampdm system architecture.}
\label{fig:rampdm-architecture}
\end{figure}

We design \rampdm, a resilient disaggregated memory system following the \ramp model.
\rampdm builds on Hydra~\cite{lee:hydra:fast:2022}, a state-of-the-art resilience mechanism for disaggregated memory systems that employs rack-level memory replication to tolerate memory-node failures. 
\rampdm extends Hydra with cross-layer memory resilience mechanisms, enabling it to tune memory protection strength, storage overhead, and performance.

\subsection{System Overview}

Figure~\ref{fig:rampdm-architecture} shows \rampdm's system architecture. 
\rampdm has three main components to support cross-layer resilience: 
(i) \emph{Resilience Manager} coordinates resilience operations during remote read/write, 
(ii) \emph{Resource Monitor} handles memory management in a memory node, and 
(iii) \emph{FlexDIMM} offers a memory module with configurable memory error protection. 
Resilience Manager and Resource Monitor are software-level components inherited from Hydra~\cite{lee:hydra:fast:2022}, which together provide the memory-replication tier.
FlexDIMM is a new hardware-level component, which provides the memory-protection tier.
Resilience Manager and Resource Monitor are extended to interact with FlexDIMM via data-plane and control-plane operations, working together to provide efficient memory resilience.
Data-plane operations for accessing data in remote memory use one-sided remote DMA (RDMA) operations (RDMA READ/WRITE), which enable fast accesses to remote memory without involving the remote node processor.
Control-plane operations for configuring remote memory use two-sided RDMA operations (RDMA SEND/RECV). 
In this architecture, each machine can function as a compute node, a memory node, or both. 

\subsection{Configurable Memory Protection}

FlexDIMM provides configurable memory error protection by reusing the memory chip failure protection bits to detect and opportunistically correct bit errors at high performance.
FlexDIMM targets random bit cell errors, as we expect these to represent the majority of memory errors because of the high RBER of high-density memory technology~\cite{zhang:pm-chipkill:micro:2018, patil:dve:isca:2021}.

We design a FlexDIMM that provides a configurable chipkill-correct protection scheme, leveraging a recent chipkill design for high-density non-volatile memory~\cite{zhang:pm-chipkill:micro:2018}. 
The module employs two protection codes: 
(i) a fixed-protection code reuses the chip failure protection bits to opportunistically correct bit errors at high performance, and 
(ii) a configurable-protection code uses long ECC codewords to correct at a configurable storage cost bit errors that are detected but uncorrected by the fixed-protection code. The configurable code 
uses a BCH(n,k,t) code for each ECC codeword of length $n=k+t(\left \lceil{log_2 (k)} \right \rceil+1)$ to correct $t$ bad bits when protecting $k$ bits of data~\cite{zhang:pm-chipkill:micro:2018}.

The FlexDIMM provides a cache-line failure probability due to DUE that is the product of two terms:
the probability that the fixed-protection code fails to correct a bit error (whose value equals to 0.018 as is taken from the original design~\cite{zhang:pm-chipkill:micro:2018}) and 
the probability that the configurable-protection code fails to correct multiple bit errors in the BCH codeword (which happens when there are at least $t$ bit errors):

\begin{equation*}
\pcdue= 0.018 \times \sum_{i=t+1}^{n}\binom{n}{i}{RBER}^i\cdot{(1-RBER)}^{n-i}
\end{equation*}

The cache-line failure probability due to DUE can be used as input to the \ramp analytical model to determine and tune the length of the BCH codeword in combination with rack-scale memory replication. 
%
The model can estimate the combined DUE rate resulting from using available replicas to correct DUEs. 
Although not shown, the cache-line failure probability due to NDE can be also calculated and used as input to the model, following the analysis of Kim and Lee~\cite{kim:undetected-error-bch:ieee-tc:1996}.

The FlexDIMM leverages existing hardware error reporting mechanisms, such as Intel Machine Check Architecture (MCA)~\cite{intel:mce:book:2024}, to raise a machine check exception (MCE) in response to uncorrectable memory errors.
For error reporting mechanisms that do not provide a mechanism for detecting write failures, like Intel MCA, the module may issue an additional read after a write to check the success of the write.
While the module, in theory, offer only error detection without correction, this would leave error correction entirely to the upper memory-replication tier. 
However, the ability to correct errors opportunistically is crucial to avoid a flood of interrupts, which could severely slowdown the system~\cite{meza:dramfailures:dsn:2015}, as corroborated in the evaluation.

The default response to an MCE in the kernel is a kernel panic. However, recent Linux kernels allow returning an error to the caller instead of crashing in response to memory-error induced MCEs~\cite{xu:nova-fortis:sosp:2017}.
After the exception, MCA registers hold information that allows the OS to identify the memory address responsible for the exception, allowing the OS to report the error to the memory-replication tier for further handling.
With this approach, after error reporting, a memory node remains operational and continues to serve memory accesses to non-failed memory regions.
Thus, in contrast to previous work where uncorrectable memory errors may crash a memory node~\cite{shan:legoos:osdi:2018} making all data stored on that node unavailable, our fine-grain failure model enables the memory node to remain operational and serve future requests, improving availability.

Propagating MCEs from the FlexDIMM, which provides the memory-protection tier, to the Resilience Manager, which provides the memory-replication tier, is challenging due to the one-sided semantics of RDMA. 
One approach would have the remote OS handle exceptions by disregarding the error, completing pending DMA requests that triggered the Memory Corrective Error (MCE), and making a callback remote procedure call (RPC) to the Resilience Manager to report the error. 
However, since one-sided operations issued by the Resilience Manager directly access memory and do not involve the remote memory node processor, this process may introduce a race condition between when the Resilience Manager receives the callback RPC and the completion of the one-sided RDMA operation. 
To address this, the FlexDIMM requires extending the RDMA NIC hardware, which handles the RDMA request, to support propagating memory errors to the Resilience Manager by piggybacking on the existing error reporting mechanism of RDMA. 
The OS kernel informs the NIC of the MCE, and the NIC reports the memory error back to the Resilience Manager as part of the response to the RDMA operation, indicating an ECC failure that requires further handling.

\subsection{Rack-scale Memory Replication}

The Resilience Manager works together with the Resource Monitor to provide the memory-replication tier. This tier uses rack-scale replication to correct memory errors that are detected but uncorrected by the memory-protection tier. 
The Resilience Manager provides a remote memory abstraction to client applications, exposing remote memory via a paging system integrated with virtual memory. 
This system transparently moves memory pages between local and remote memory using one-sided RDMA operations, enabling applications to access remote memory without any changes to the application code. 
The Resilience Manager runs in kernel mode and runs on every client (compute node) that needs access to remote memory.  
The Resource Monitor manages memory on a memory node and makes it available to the remote Resilience Manager. 
The Resource Monitor runs in user space and only participates in control-plane activities, responding to requests from Resilience Manager for allocating physical pages and configuring page protection settings.

The Resilience Manager handles all aspects of redundancy. 
For each virtual page, it ensures that the page is stored on multiple physical-page replicas, distributed across different nodes within the rack, to enhance availability. 
The Resilience Manager allocates physical pages through requests made to the Resource Monitor and 
and communicates with the Resource Monitor to configure the memory protection scheme and strength of each physical page to meet a target UBER and SDC rate, according to application requirements.
The Resilience Manager tracks and blacklists failed memory regions to avoid mapping replicas to regions with known errors. When the Resilience Manager encounters a data access error (DUE), it attempts to correct the memory error using another replica.

\subsection{Putting it all together}

Applications define their required level of resilience, which is communicated to the Resilience Manager. The Resilience Manager, in turn, ensures the appropriate redundancy is in place to meet these resilience requirements.
Applications access remote memory through virtual memory. 
When an application attempts to access a virtual memory address that is not backed by a physical page in local memory, a page fault occurs. This triggers remote paging to retrieve the corresponding page from remote memory over the network using one-sided RDMA operations. 
The memory controller of the remote memory node processing the RDMA request uses the page protection information to identify which protection technique to use for any given memory access.
If a memory access triggers a DUE, the memory node propagates it back to the Resilience Manager as an RDMA error, which is then handled by the Resilience Manager using other remote replicas.
\section{Methodology}
We outline the methodology used to evaluate \rampdm.

\mypar{Implementation}
We base our \rampdm prototype on the publicly available Hydra remote memory system~\cite{lee:hydra:github}, which we extend to handle synthetic memory faults as described below. 
We model the performance overhead of the chipkill design by throttling memory bandwidth to 8/9 of the available memory bandwidth utilizing the memory controller's thermal control registers~\cite{volos:quartz:middleware:2015} to account for the additional overhead to read 8B parity bits for every 64B block~\cite{zhang:pm-chipkill:micro:2018}.

\mypar{Fault Injection}
We model a target DUE rate through synthetic memory fault injection. Our aim is to inject faults that cause Machine Check Exceptions (MCEs) to exercise the complete hardware and software path for handling such interrupts. 
Although server systems and the Linux kernel provide various methods for injecting MCEs for testing, we find none of these methods suitable for our needs. 
Software-level simulation methods~\cite{linux:madvise-hwpoisson:2024, linux:mce-inject:2024} utilize virtual memory protection to poison the pages in a specified memory range, handling subsequent references to those pages like a hardware memory corruption. Hence, these methods do not simulate full MCE handling on the platform level.
In contrast, firmware-level simulation methods~\cite{linux:einj:2024} trigger actual MCEs, fully exercising MCE handling. Despite successfully triggering MCEs synchronously through direct API calls, we were unable to trigger them asynchronously through one-sided RDMA operations.
To address this gap, we developed a custom synthetic fault injection framework.
We extend the Resilience Manager with a fault injector that injects faults at a configurable rate. 
When a fault needs to be injected, the injector performs a two-sided send operation to the resilience monitor, triggering a synchronous MCE~\cite{linux:einj:2024}, rather than performing an one-sided RDMA operation.
This approach captures both the communication overhead of a one-sided RDMA operation that would fail due to a memory error and the overhead associated with handling the error-induced MCE.

\mypar{Cluster Setup}
We evaluate \rampdm on a cluster consisting of 12 machines interconnected via a 56 Gbps InfiniBand network, hosted on CloudLab~\cite{cloudlab:hardware}.
Each machine is equipped with a Xeon E5-2450 processor featuring 8 cores and 16 GB of physical memory. 
Each machine runs Linux kernel 4.4 with Mellanox OFED 4.1.
We run a single Resilience Manager on a single machine with a remote memory capacity of 1GB. 
The manager pages and replicates to remote memory allocated from six other machines, each running its own Resource Monitor.

\mypar{Workload Setup}
We evaluate \rampdm using Memcached~\cite{memcached}, a lightweight in-memory key-value store that is widely deployed as a distributed caching service to accelerate user-facing applications with stringent latency requirements~\cite{nishtala:memcached:nsdi:2013, yang:twemcache:osdi:2020}.
We run a single Memcached server process inside an LXC container on the machine running the Resilience Manager. The container limits the local memory capacity to a configurable amount, effectively forcing Memcached satisfy any additional memory requirements by paging to remote memory through the Resilience Manager.
We drive Memcached using an extended version of the Mutilate load generator~\cite{leverich:mutilate:eurosys:2014} configured to recreate the ETC workload from Facebook~\cite{berk:facebook-kv-workload:sigmetrics:2012}, using one master and two workload-generator clients, each running on a separate machine. 
We populate Memcached with 3 million records, for a total footprint of 1GB memory.
We limit local memory to 25\% of the total footprint.

\newcommand{\chipkill}{\emph{Chipkill}\xspace}
\newcommand{\chipkillrep}{\emph{Chipkill-REP}\xspace}
\newcommand{\chipkillec}{\emph{Chipkill-EC}\xspace}

\mypar{Configurations}
We study two protection schemes: 
(i) \chipkill, a baseline scheme that uses chipkill alone without rack-scale redundancy, 
(ii) \chipkillrep, a rack-scale redundancy scheme that combines rack-scale replication with chipkill, and 
(iii) \chipkillec, a rack-scale redundancy scheme that combines rack-scale erasure coding with chipkill. 
For the two schemes, we choose parameters so that they can both tolerate up to two replica failures (following standard practice), that is N=3 for replication and N=6 and K=4 for erasure coding.
We vary storage overhead by varying the protection strength of the BCH code to protect individual replica blocks.
We vary strength by varying the number of $t$ bit errors that can be corrected by the BCH code.
We assume uniform access to all logical blocks and that all physical blocks are equally vulnerable to memory errors.

\mypar{Memory Technology}
Unless otherwise specified, we assume a memory technology with $RBER=2\times10^{-4}$, as in~\cite{zhang:pm-chipkill:micro:2018}.

\section{Evaluation}

\subsection{Performance Overhead due to Error Handling}

\begin{figure}[tb]
\centering
\includegraphics[width=3.25in]{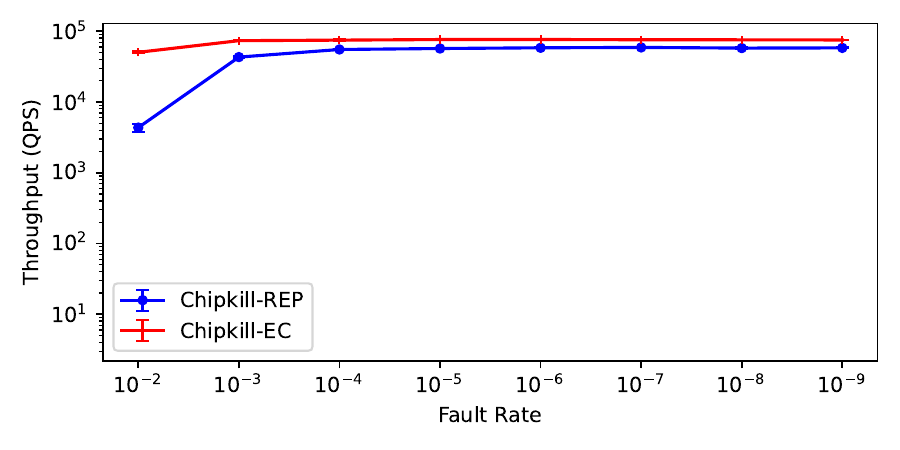}
\caption{Memcached throughput as completed queries per second (QPS) for different DUE fault rates.}
\label{fig:throughput-with-failures}
\end{figure}

\begin{figure}[tb]
\centering
\includegraphics[width=3.25in]{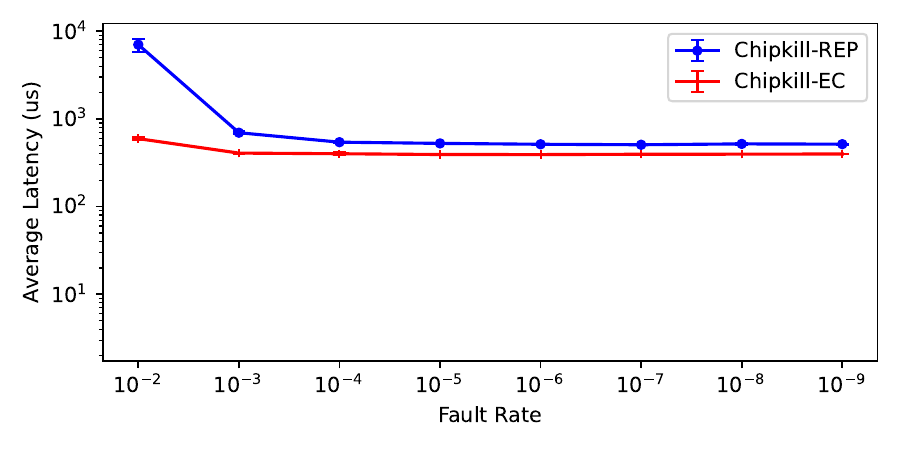}
\includegraphics[width=3.25in]{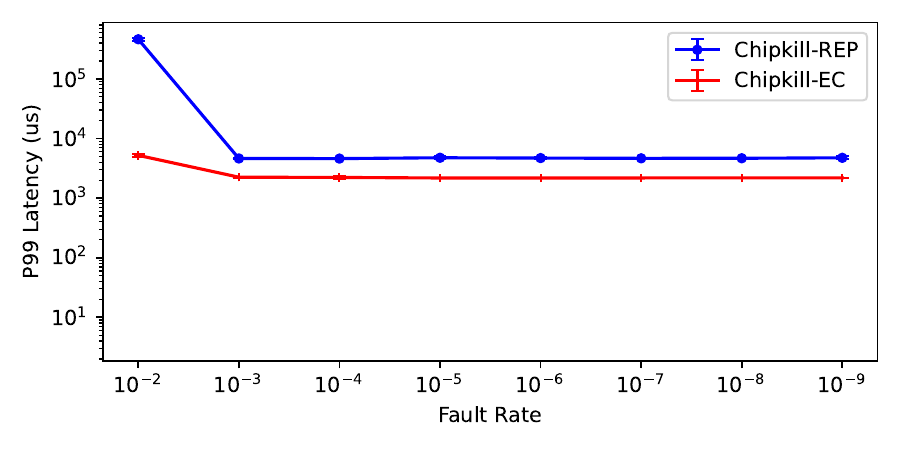}
\caption{Memcached average and P99 response latency for different DUE fault rates.}
\label{fig:latency-with-failures}
\end{figure}

Figures~\ref{fig:throughput-with-failures} and \ref{fig:latency-with-failures} illustrate the throughput and response latency performance of the two rack-scale configurations, \chipkillrep and \chipkillec, with DUE fault rates of individual replicas ranging from $10^{-2}$ to $10^{-9}$. 
Performance overhead includes the impact of replication and erasure coding.
We find that the chipkill bandwidth overhead does not have impact on end-to-end performance as network bandwidth (56 Gbps = 7GB/s) is the bottleneck, despite throttling memory bandwidth to 19GB/s to account for the chipkill overhead.

As predicted by the analytical model, at low DUE rates, accessing additional replicas is infrequent, leading to negligible performance overhead. The graphs demonstrate that this threshold is reached at a DUE rate of approximately $10^{-5}$, indicating that the increase in MCE rate due to the weaker protection strength of each replica does not significantly impact performance until the DUE rate becomes considerably higher. 
To better understand the performance overhead introduced by an MCE, we manually inject a synchronous MCE~\cite{linux:einj:2024} and measure the time required to handle it. Our results show that the average MCE handling latency is 200 $\mu$s, with occasional spikes up to 1 second. This latency helps explain why, under aggressive fault rates, the performance impact can become so pronounced.

Interestingly, the \chipkillec configuration, which uses erasure coding, outperforms \chipkillrep, despite the added computational complexity of erasure coding. This can be attributed to the memory-bound nature of Memcached, which leaves the Resilience Manager with ample computational resources to handle the erasure coding process. Furthermore, \chipkillec’s erasure coding approach enables multiple small accesses to different replicas in parallel, rather than a single larger page access to a single replica. This parallelism better utilizes available bandwidth across replicas and helps avoid potential bottlenecks at any one replica.

\subsection{Storage Savings}
We use the analytical model to study availability and storage-saving trade-offs for the two rack-scale configurations, \chipkillrep and \chipkillec, against the baseline \chipkill configuration.

\begin{figure}[tb]
\centering
\includegraphics[width=3.5in]{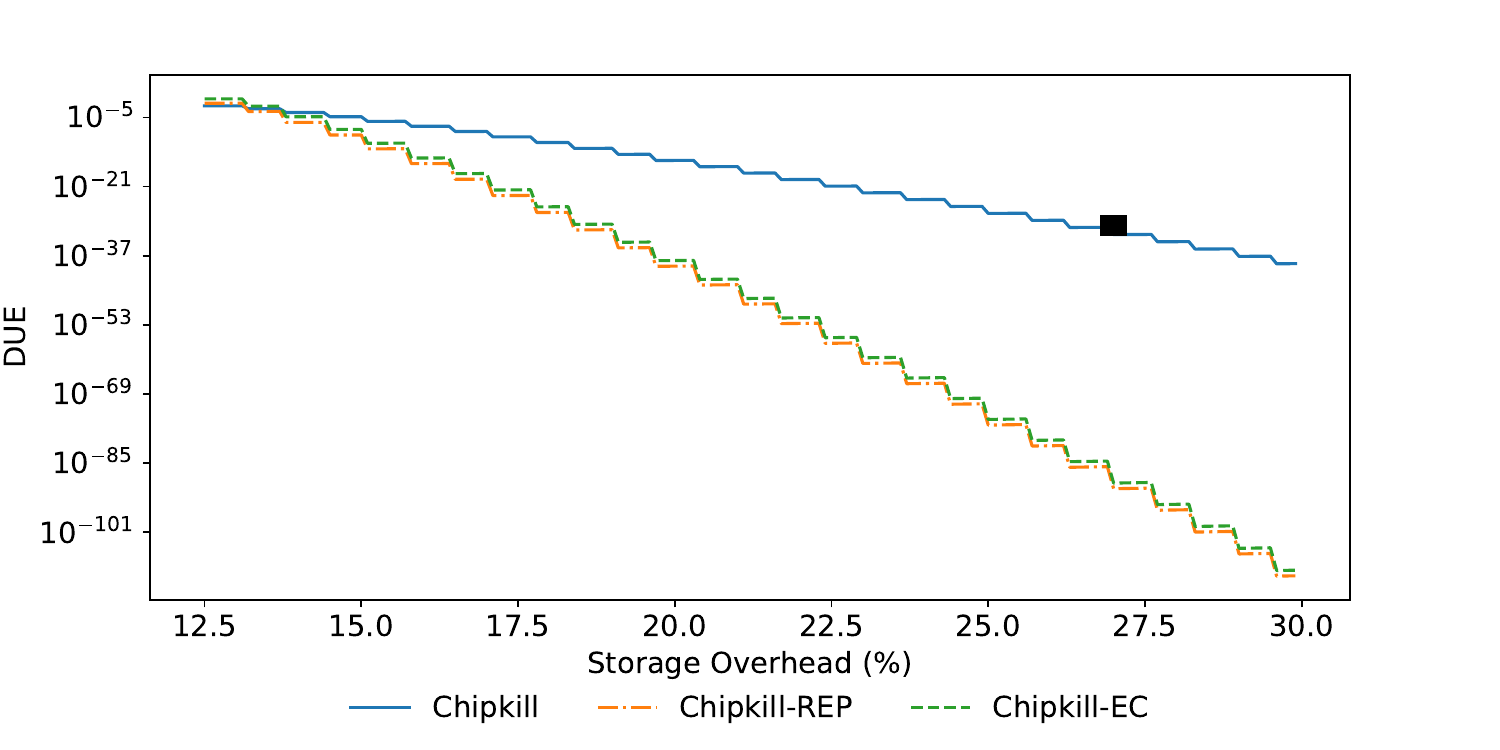}
\includegraphics[width=3.5in]{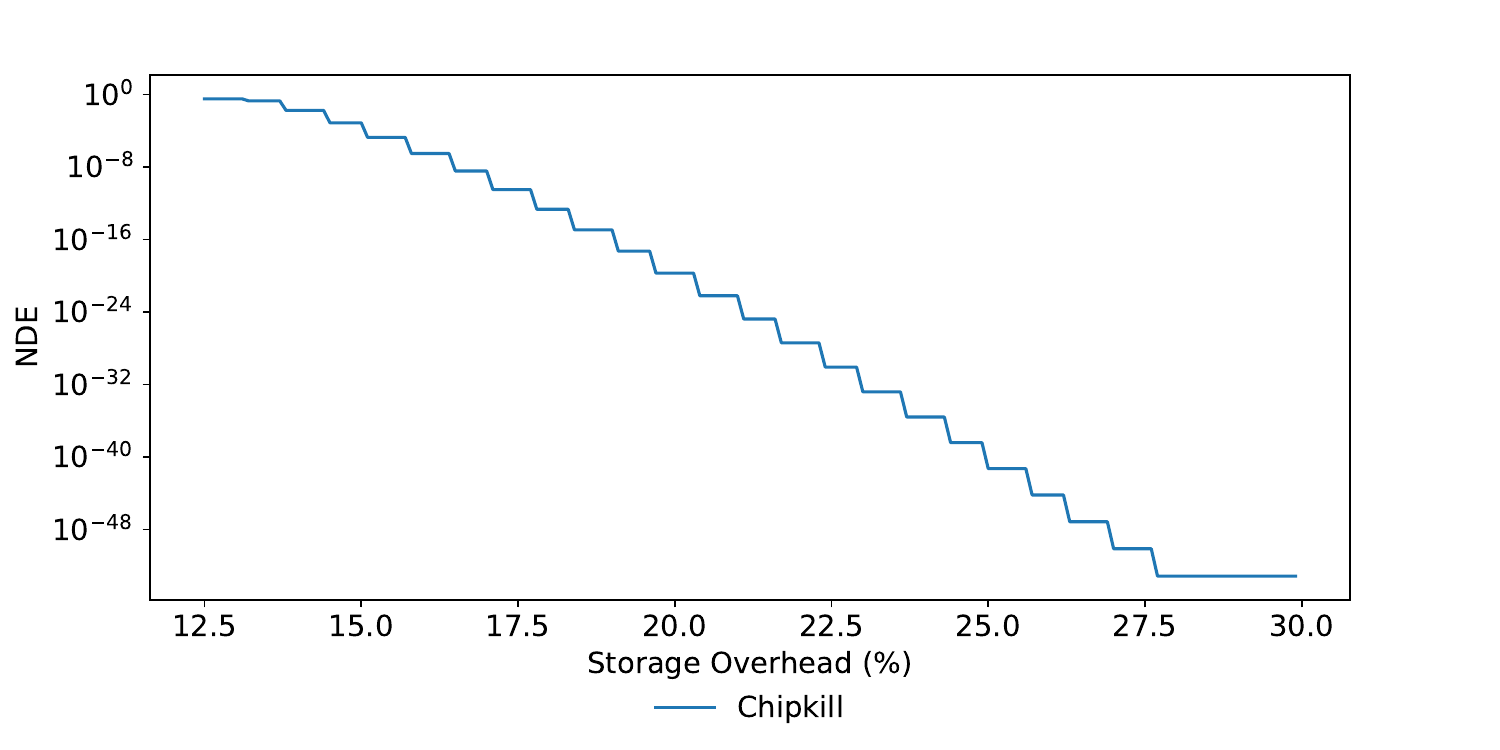}
\caption{DUE and NDE versus storage overhead for different chipkill protection schemes.
The solid rectangle (in the top figure) marks the DUE and storage overhead of the original chipkill design~\cite{zhang:pm-chipkill:micro:2018}.
NDE is shown only for baseline chipkill as it is independent of replication and identical for all chipkill schemes.
}
\label{fig:storage-overhead-due}
\end{figure}

\begin{figure}[tb]
\centering
\includegraphics[width=3.5in]{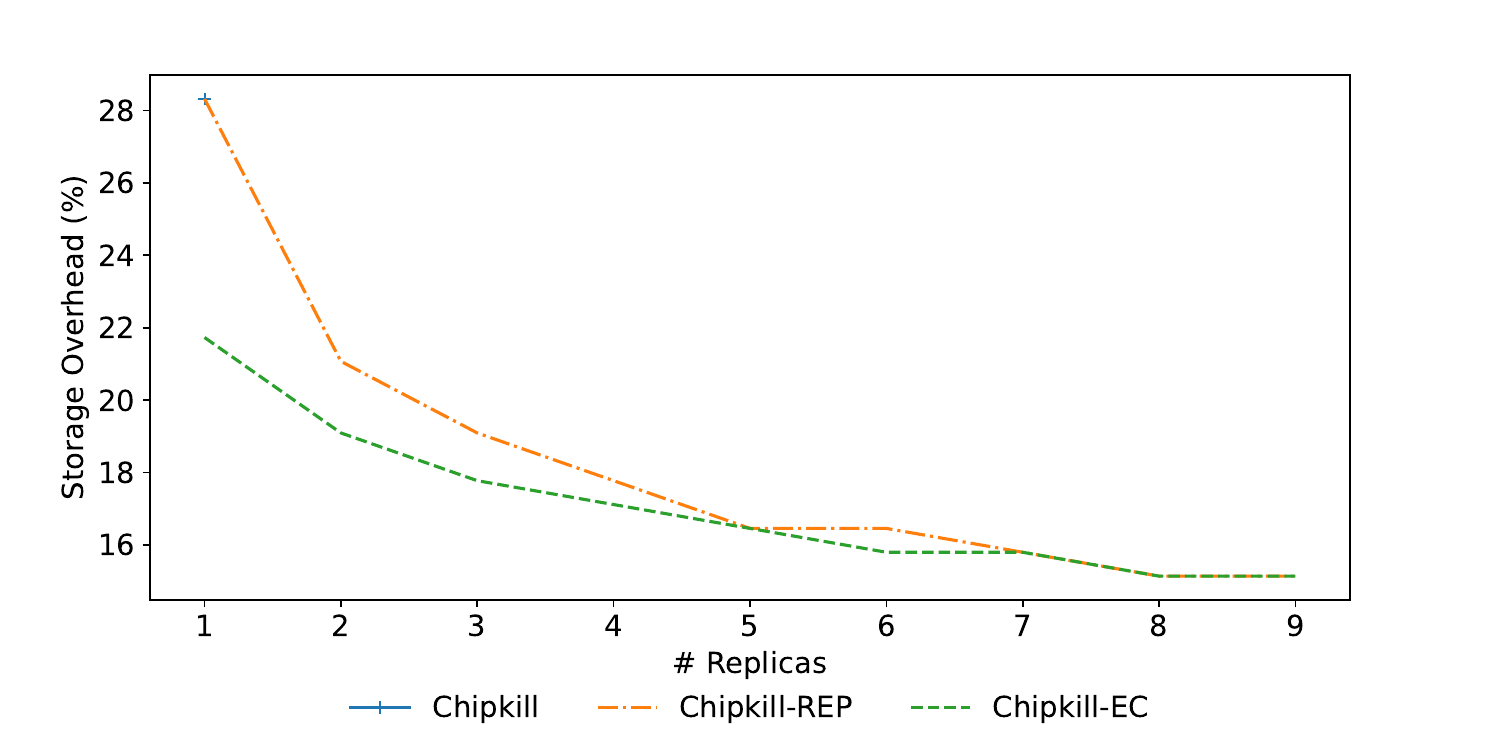}
\caption{Storage overhead versus number of replicas for different chipkill protection schemes.
}
\label{fig:storage-overhead-replicas}
\end{figure}

\begin{figure}[tb]
\centering
\includegraphics[width=3.5in]{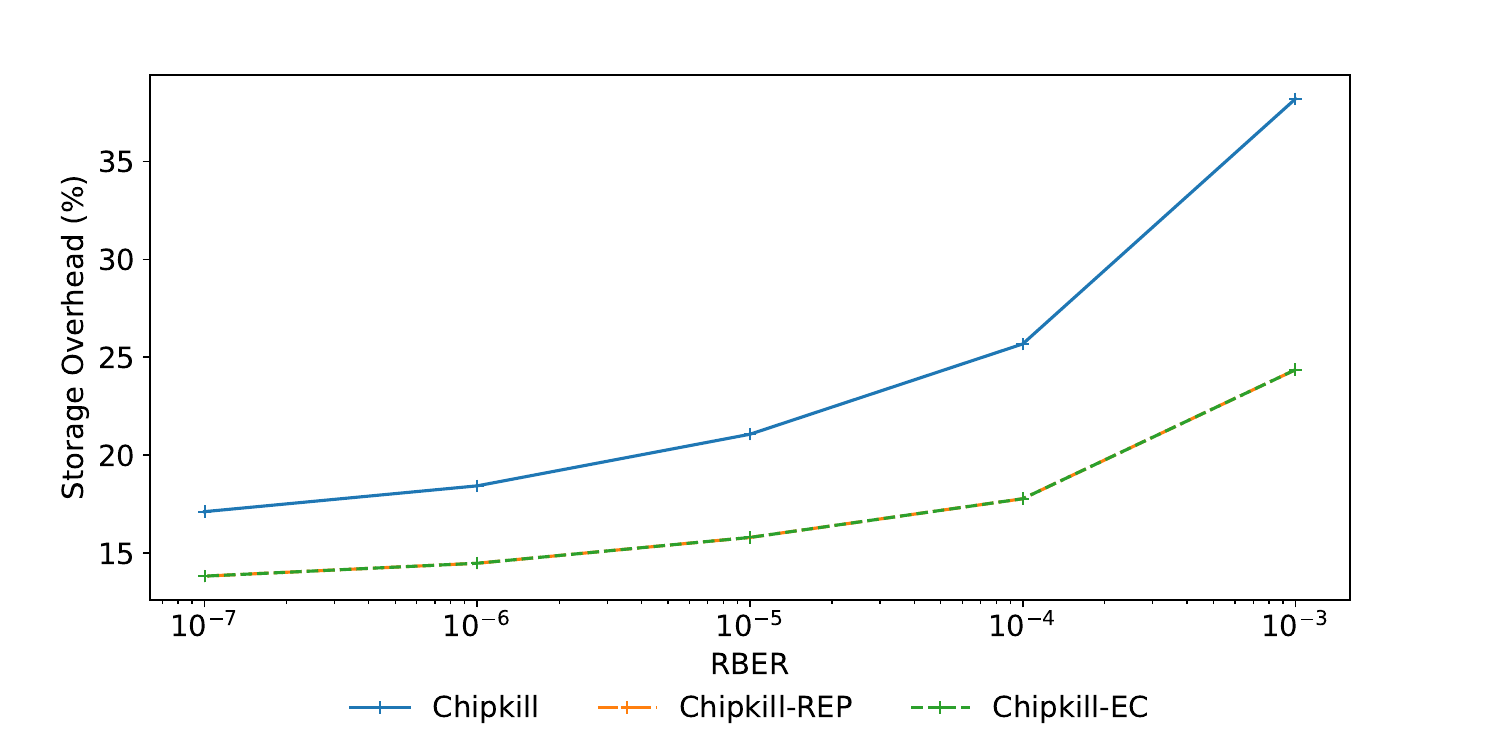}
\caption{Storage overhead versus RBER for different chipkill protection schemes.
}
\label{fig:storage-overhead-rber}
\end{figure}

Figure~\ref{fig:storage-overhead-due} plots combined DUE rate and baseline NDE rate of individual physical blocks as a function of storage overhead.
For each replication scheme, the storage overhead is calculated over a corresponding baseline that employs the same replication scheme but without chipkill protection. 
For \chipkillrep, we use a physical block size equal to the virtual page size, that is 4KB.
For \chipkillec, we use a physical block size equal to 1KB ($\frac{4KB}{4}$).
We observe that both replication and erasure coding can achieve the same level of DUE as the original ckipkill design ($\sim10^{-33}$ DUE rate), albeit at about $9\%$ less overhead.
\revisionhighlight{
For a target SDC rate of $10^{-22}$~\cite{zhang:pm-chipkill:micro:2018}, we need to provision an extra $2.4\%$ overhead, bringing the storage  savings down to $6.6\%$.
}

Figure~\ref{fig:storage-overhead-replicas} plots the storage overhead sustained to achieve the same level of DUE as the original chipkill design as we vary the number of replicas.
We observe diminishing returns in storage savings as we increase the number of replicas, suggesting that \ramp could be more beneficial with low replication factors. 

Finally, Figure~\ref{fig:storage-overhead-rber} plots the storage overhead sustained to achieve the same level of DUE as the original chipkill design as we vary the raw bit error rate (RBER) of the memory technology.
We observe growing returns in storage savings as we increase the RBER, suggesting that \ramp could be more beneficial with higher-density memories exhibiting higher memory error rates. 

Overall, these results confirm our main hypothesis: by weakening the protection of each individual replica, we can lower the storage overhead while we can rely on the combined protection conferred by multiple replicas to meet a stronger protection target.

\section{Related Work}

\mypar{Tolerating Memory Errors}
Several approaches have been proposed to protect data from memory errors in NVM. One method optimizes chipkill-correct to address bit errors in high-density NVM with high random raw bit error rates (RBER)~\cite{zhang:pm-chipkill:micro:2018}. Pangolin~\cite{zhang:pangolin:atc:2019} and NOVA-Fortis~\cite{xu:nova-fortis:sosp:2017} use checksums and parity to protect application and file system data, respectively, from media errors in NVM. TVARAK~\cite{kateja:tvarak:isca:2020} offloads the management of checksums and parity to a hardware controller co-located with the last-level cache, reducing software overhead while protecting against memory errors. In contrast to our work, all of these solutions focus on error correction within a single memory node and do not ensure high availability in cases where a memory node becomes completely unavailable.

\mypar{Fault-tolerant Disaggregated Memory}
With thousands of interconnected compute and memory devices, failures become commonplace, thus making fault tolerance a critical property of any practical approach for disaggregated memory. Failures, if not addressed properly, may result in data loss and force tasks to stop and restart. Although most prior frameworks for disaggregated memory lack fault tolerance, recent proposals can tolerate memory node failures through replication or erasure coding~\cite{zhou:carbink:osdi:2022, lee:hydra:fast:2022}. Our work complements these approaches by providing an efficient cross-layer resilience method for tolerating memory errors in high-density NVM.

\mypar{Cross-layer Resilience}
DIRECT~\cite{tai:flash-uber:atc:2019} is an application-level cross-layer resilience technique that leverages replication in distributed storage systems, such as key-value stores, to efficiently handle uncorrectable errors in flash storage. 
In contrast, our work operates at the memory interface and utilizes both distributed erasure coding and replication to provide enhanced redundancy for improved memory error resilience.

\section{Conclusion}
We introduced \ramp, a framework for designing and tuning computing systems combining memory replication with memory protection to tolerate memory errors efficiently.
We demonstrate the utility of the \ramp framework by applying it to a state-of-the-art resilient disaggregated memory design that utilizes memory replication for availability, enabling a reduction in storage overhead for memory protection without compromising overall resilience.

A current limiting factor in aggressively relaxing memory protection for higher storage savings is the risk of increasing NDEs, where silently corrupted data might escape correction by the memory-replication tier. Future work could explore application-level checksumming to detect silent data corruption in the memory-replication tier. 


Overall, our results are promising, demonstrating that \ramp can facilitate the cost-optimization of two-tier memory resilience schemes. Our case study shows that relaxing memory protection strength can yield significant storage savings without compromising overall resilience, all while maintaining minimal performance overhead.

\vspace{-0.1cm}
\ifCLASSOPTIONcompsoc
  \section*{Acknowledgments}
\else
  \section*{Acknowledgment}
\fi

This project has received funding from the European Union’s Horizon 2020 research and innovation programme under the Marie Skłodowska-Curie grant agreement No 101029391.

\bibliographystyle{plain}
\bibliography{main}

\end{document}